\documentclass[%
 reprint,
 aps,
 prl,
 amsmath,amssymb,
 superscriptaddress,
 nofootinbib,
 floatfix,
]{revtex4-2}

\usepackage{graphicx}
\usepackage{booktabs}
\usepackage{bm}
\usepackage[dvipsnames]{xcolor}
\usepackage[colorlinks=true,citecolor=blue,urlcolor=blue,linkcolor=red]{hyperref}
\usepackage{orcidlink}

\makeatletter
\let\frontmatter@footnote@produce\frontmatter@footnote@produce@footnote
\makeatother

\newcommand{\LCDM}{$\Lambda$CDM}

\newcommand{\iB}{b}            
\newcommand{\iM}{m}            
\newcommand{\iT}{t}            
\newcommand{\Rfs}{f_{\rm fs}}
\newcommand{\fps}{f_{\rm ps}}
\newcommand{\GamT}{\Gamma_T}   
\newcommand{\Gamo}{\Gamma_0}   
\newcommand{\YP}{Y_P}
\newcommand{\Om}{\Omega_m}
\newcommand{\xe}{x_e}

\begin{document}

\title{Four consistent estimates of $H_0$ from a \LCDM\ inference network given CMB data}

\author{Gabriel P. Lynch\,\orcidlink{0009-0004-3143-1708}}
\email{gplynch@ucdavis.edu}
\affiliation{Department of Physics and Astronomy, University of California, Davis, California, USA 95616}

\author{Lloyd Knox}
\affiliation{Department of Physics and Astronomy, University of California, Davis, California, USA 95616}

\date{\today}

\begin{abstract}
We organize inferences from cosmic microwave background (CMB) data, of the current rate of expansion, $H_0$, into a network whose structure is informed by a scaling symmetry of the Einstein-Boltzmann evolution equations. There are four distinct routes through this network from a calibrating anchor to $H_0$. Given key assumptions of the \LCDM\ model, they are mutually consistent. This consistency is a significant feature of the \LCDM\ model's compatibility with CMB data. The redundancy in $H_0$ determinations illuminates a hurdle to resolving the $H_0$ tension with alternative cosmological models. 
\end{abstract}

\maketitle

\textit{Introduction.---} The rate of expansion of space in the current era, $H_0$, as inferred via the classical distance ladder (CDL)~\citep{H0DN:2025lyy,Breuval:2024lsv}, is discrepant at greater than 6$\sigma$ with the value given by the standard cosmological model, once its parameters are constrained by observations of the cosmic microwave background (CMB)~\citep{Planck2018,AtacamaCosmologyTelescope:2025blo,SPT-3G:2025bzu}.\footnote{For a dissenting view of the significance of this discrepancy, see \citet{Freedman:2024eph}.} This discrepancy, often referred to as the `Hubble tension', has received much attention from observers, attempting to find or constrain systematic errors \citep{Freedman:2024eph,H0DN:2025lyy}, and from theorists, building cosmological models that are consistent with the CDL results, the CMB, and other cosmological data~\citep{CosmoVerseNetwork:2025alb,Schoneberg:2026eys}. Despite these efforts, the tension persists.

In this {\it Letter} we focus on the cosmological-model-dependent path to $H_0$, that is, its inference from CMB data given a model. We introduce a general framework for understanding such inferences, which organizes them into a network with a structure informed by a scaling symmetry of the relevant evolution equations. The symmetry was first identified by \citet{Cyr-Racine:2021oal}, who used it to build a model capable of resolving the Hubble tension,\footnote{We return to this model, and the difficulty it faces, in the Discussion.} and later to explain constraints on light relics~\cite{Ge:2022qws}.

The network can be thought of as a theoretical analogue of an observational network, the `Local Distance Network' (LDN)~\cite{H0DN:2025lyy}. We use the theoretical network to identify four distinct routes to $H_0$ given CMB data, each relying on a different modeling assumption and on different physical effects. Taking the relevant \LCDM\ assumption for each, we infer four values of $H_0$ and find them to be consistent with one another.

We also briefly discuss i) a $3\%$ determination of the baryon fraction from its gravitational signature in the CMB alone, ii) how the existence of four consistent routes presents a challenge to cosmological-model-based solutions to the Hubble tension, iii) use of the network for categorizing such solutions, and iv) a possible origin of the tendency for these solutions to prefer a baryon density higher than that inferred from big bang nucleosynthesis (BBN)~\cite{Giovanetti:2026aku}.

\textit{Scaling symmetry revisited.---} We first revisit the scaling symmetry introduced by \citet{Cyr-Racine:2021oal}, to highlight the main consequence relevant for the discussion below: that the CMB power spectra, on their own, determine only dimensionless ratios of densities and rates. Linearized about a homogeneous background, the Einstein--Boltzmann (EB) equations have the same basic form for all the models we consider. A model and its parameters enter only through the following inputs: the free-fall rates of all species $\sqrt{G \rho_i(a)}$, which determine the expansion rate\footnote{Here and throughout the text, we assume zero mean spatial curvature.}; the rates of local interactions; and dimensionless functions such as equation-of-state parameters. Let $g_\lambda$ rescale every rate and wavenumber $k$ by $\lambda$, leaving the dimensionless functions unchanged. Because the constants of nature enter only in combination with one of these inputs, $g_\lambda$ acts on the equations as a change of units, and the dimensionless anisotropies are unchanged~\cite{Cyr-Racine:2021oal}:
\begin{equation}
  {\rm EB}\circ g_\lambda={\rm EB},
  \label{eq:symmetry}
\end{equation}
with ${\rm EB}$ denoting the Einstein--Boltzmann system mapping these inputs to observables. The primordial spectrum must also transform (End Matter). Its parameters set the initial conditions, and we marginalize over them, and over the optical depth to reionization, throughout.

The symmetry is broken by any independently fixed scale, such as a particle mass or the atomic rates of recombination; however, the symmetry-breaking effects of each of these are mild~\cite{Cyr-Racine:2021oal}. We therefore take the neutrinos to be massless and hold the ionization history $\xe(a)$ fixed (see End Matter for implementation details), so that the symmetry is exact. Cosmologies related by $g_\lambda$ then have identical spectra, which carry no information about the absolute scale of the rates, the expansion rate included.

\textit{Amplitudes and invariants.---} An analysis of CMB data, assuming the \LCDM\ model, nevertheless returns an estimate of $H_0$. This is because an additional measurement, the FIRAS determination of the CMB temperature \citep{FIRAS1,Fixsen:1996nj}, provides us with the photon energy density $\omega_\gamma$. Together with certain assumptions of the model linking the photon density to the rates mentioned above, the overall scale of these rates, and therefore the expansion rate, is also fixed by this measurement. Our aim is to identify assumptions in \LCDM\ that make these links, so that we may relax all but one at a time, in order to assess the consistency of the $H_0$ determinations following from each. To do so, we first identify the relevant inputs to the EB equations in \LCDM, together with the dimensionful constants that specify them (hereafter, \emph{amplitudes}), and then determine which of these are computed from $\omega_\gamma$ in \LCDM.

Five components source gravity. Because gravity couples only to stress-energy, we label them not by species but by how their stress-energy evolves before recombination: radiation that free-streams, radiation that does not, matter that is pressure supported, matter that is not, and the cosmological constant. The gravitational response to radiation that does not free-stream cannot, for example, tell us that this radiation is all photons. The usual identification of the first three with the neutrinos, the photons, and the baryons is thus an assumption of the model, to which we return below. Whatever its composition, each of the five has its own free-fall rate. The density of each scales as a fixed power of $a$, so its amplitude is its present density, $\omega_X\propto\rho_{X,0}$: we write $\omega_{fs}$, $\omega_{fld}$, $\omega_{ps}$, $\omega_c$ and $\omega_\Lambda$ for the five in the order listed, and $\omega_r=\omega_{fs}+\omega_{fld}$ and $\omega_m=\omega_{ps}+\omega_c$ for all of the radiation and all of the matter. The one non-gravitational interaction rate is that of Thomson scattering, $\GamT(a)=\Gamo\,\xe(a)\,a^{-3}$, with amplitude $\Gamo$. We will refer to these six quantities as rate amplitudes.

\begin{table}[t]
\centering
\footnotesize
\setlength{\tabcolsep}{2pt}
\begin{tabular}{@{}llll@{}}
\toprule
Invariant & Definition & Characteristic effect & Value \\
\midrule
$\iB$              & $\omega_b/\omega_\gamma$   & odd/even peak heights     & $898.6\pm6.1$ \\
$\iM$              & $\omega_m/\omega_r$        & radiation driving         & $3415\pm36$ \\
$\Rfs$             & $\omega_{fs}/\omega_r$     & phase shift of the peaks  & $0.398\pm0.030$ \\
$\iT$              & $\Gamo/\sqrt{\omega_r}$    & damping tail              & $2.653\pm0.031$ \\
$\fps$             & $\omega_{ps}/\omega_m$     & potential evolution       & $0.1580\pm0.0047$ \\
$100\,\theta_s$    & $100\,r_s/D_A$             & peak spacing              & $1.04178\pm0.00073$ \\
\bottomrule
\end{tabular}
\caption{The six invariants, an effect in the spectra characteristic of each (see \citep{Planck:2016tof} for a review), and the mean and standard deviation of each from primary CMB data (see Results). $\Gamo$ is in units of $\sigma_T\rho_{100}/m_p$, with $\rho_{100}$ the critical density for $h=1$. Given the other invariants, $\theta_s$ is equivalent to $\Om=\omega_m/h^2$, which is constrained to $0.3164\pm0.0084$. For the effect of $\fps$, see Fig.~3 of \citet{Ge:2022qws}. }
\label{tab:invariants}
\end{table}

The densities of the photons and baryons themselves, with present values $\omega_\gamma$ and $\omega_b$, enter the EB equations in one further place: in the photon--baryon collision term, as the ratio of their inertias. Together with the six rates just introduced, this ratio completes the list of inputs. It is proportional to $\rho_b/\rho_\gamma=(\omega_b/\omega_\gamma)\,a$, and is specified by the single number, $\iB\equiv\omega_b/\omega_\gamma$, which the spectra determine (Table~\ref{tab:invariants}). Once $\iB$ is known, we can consider the FIRAS measurement as providing both $\omega_\gamma$ and $\omega_b$, so that rate amplitudes which in \LCDM\ are computed from $\omega_b$ likewise receive the absolute calibration.

In \LCDM\ $\omega_c$ and $\omega_\Lambda$ are free parameters while $\omega_{fs}$, $\omega_{fld}$, $\omega_{ps}$, and $\Gamma_0$ can be computed from $\omega_\gamma$ and $\omega_b$, by assuming: (i) that the radiation is the photons together with neutrinos whose density follows from the photon temperature (and standard assumptions of the thermal history), which fixes $\omega_r$; (ii) that the pressure-supported matter is the baryons, $\omega_{ps}=\omega_b$; (iii) that the free-streaming radiation is those same neutrinos, which fixes $\omega_{fs}$; and (iv) that the Thomson rate follows from $\omega_b$, given the helium fraction predicted by standard BBN, $\YP=\YP^{\rm BBN}(\omega_b)$, and laboratory values of $\sigma_T$ and $m_p$. These four are the assumptions that we relax below. A different division of the inputs, such as one that also freed the ionization history or admitted additional components with different equations of state, sound speeds, or viscosities, would give a different set of assumptions to be relaxed.

With these assumptions relaxed, the FIRAS calibration no longer reaches the rate amplitudes, and all that is known of them is what the spectra alone can determine. The symmetry of Eq.~\eqref{eq:symmetry} states that the spectra do not change when all six rates are rescaled together, so the spectra are sensitive only to the sizes of the rates relative to one another. Measured in units of any one of them, the other five are five pure numbers. These five numbers and $\iB$, which is already a pure number, are the six quantities that the spectra constrain. They are unchanged by $g_\lambda$, so we hereafter call them \emph{invariants}. Any function of invariants is itself an invariant, so the six may be reparametrized freely. Table~\ref{tab:invariants} lists our choice. For the one involving the cosmological constant we list the angular size of the sound horizon, $\theta_s$. However, given the other invariants it is equivalent to the matter fraction today, $\Om=\omega_m/h^2$, where $h^2=\omega_m+\omega_r+\omega_\Lambda$ and $H_0=100\,h$\,km\,s$^{-1}$\,Mpc$^{-1}$, so we use the two interchangeably. Each set of values of the invariants corresponds to a family of universes with identical spectra, differing only in the scale of the rates.

\begin{figure}[t]
  \includegraphics{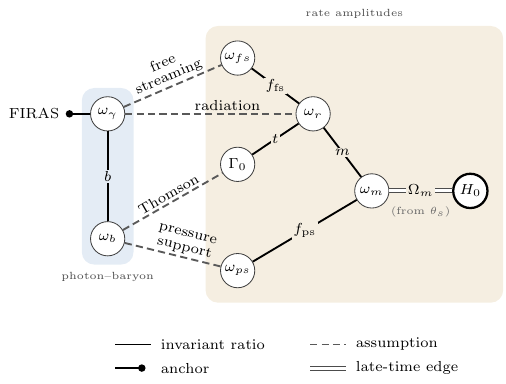}
    \caption{The \LCDM\ inference network. Nodes are $\omega_\gamma$, $\omega_b$ and the rate amplitudes, with the totals $\omega_r$ and $\omega_m$ in place of $\omega_{fld}$ and $\omega_c$, and $H_0$ in place of $\omega_\Lambda$. Solid edges are the invariants of Table~\ref{tab:invariants}, which join the nodes into two pieces. Dashed edges are assumptions (i)--(iv) of Table~\ref{tab:routes}. The dot marks the FIRAS anchor, and the late-time edge $\Om$ is obtained from $\theta_s$.}
  \label{fig:network}
\end{figure}

\textit{Inference network.---} Figure~\ref{fig:network} collects the amplitudes, the invariants and the four assumptions into a network structure. Its nodes are $\omega_\gamma$, $\omega_b$ and the rate amplitudes, with $H_0$ standing in for $\omega_\Lambda$. Each solid edge is an invariant, which converts between the nodes it connects. No invariant connects a rate amplitude to $\omega_\gamma$ or $\omega_b$, so the solid edges form two disconnected pieces: the photon--baryon densities, which FIRAS anchors, and the rate amplitudes, which can be rescaled under the symmetry. Each dashed edge is one of the four assumptions and joins the two pieces. Because all of the rate amplitudes are joined by solid edges, knowing any one of them fixes the rest, so a single assumption suffices to carry the FIRAS calibration to all of them, $H_0$ included.

Adopted individually, each assumption defines a route from the FIRAS anchor to $H_0$, which we label by the rate amplitude that the assumption fixes (Table~\ref{tab:routes}). A route crosses to the rate amplitudes along its dashed edge, follows solid edges to $\omega_m$, and ends on the edge from $\omega_m$ to $H_0$, which we call the late-time edge: $h^2=\omega_m/\Om$, with $\Om$ obtained from $\theta_s$. The pressure-support route, for example, runs from $\omega_\gamma$ along $\iB$ to $\omega_b$, across assumption (ii) to $\omega_{ps}$, and along $\fps$ to $\omega_m$, so that $\omega_m=\iB\,\omega_\gamma/\fps$. The routes differ in the assumption they cross and the invariants they use, but all four share the late-time edge.

The radiation route corresponds to the familiar account of how the CMB determines $H_0$ with a standard ruler. As described in \citet{Knox:2019rjx}, the sound horizon is first calibrated using the baryon and matter densities, the first obtained from $\iB$ and the photon temperature, the second from the matter-to-radiation ratio $\iM$. The latter is converted to the dimensionful $\omega_m$ after determining the total radiation density from the photon temperature and the neutrino content, which is assumption (i). The ruler is then applied, with $\theta_s$ giving the distance to last scattering and hence $H_0$, which is the late-time edge. In actuality, \LCDM\ makes all four assumptions at once, so its inferred $\omega_m$ and $H_0$ are informed by all four routes together. What such an analysis does not reveal, however, is whether these routes would agree if considered one at a time.

\begin{table}[t]
\centering
\footnotesize
\setlength{\tabcolsep}{2pt}
\begin{tabular}{@{}llll@{}}
\toprule
Route & Assumption & $\omega_m/\omega_\gamma$ & $H_0$ \\
\midrule
Radiation        & $\omega_r=(1+a_\nu N_{\rm eff})\,\omega_\gamma$ & $(1+a_\nu N_{\rm eff})\,\iM$ & $67.20\pm0.57$ \\
\begin{tabular}[c]{@{}l@{}}Pressure\\support\end{tabular} & $\omega_{ps}=\omega_b$ & $\iB/\fps$ & $66.7\pm1.6$ \\
\begin{tabular}[c]{@{}l@{}}Free\\streaming\end{tabular} & $\omega_{fs}=a_\nu N_{fs}\,\omega_\gamma$ & $a_\nu N_{fs}\,\iM/\Rfs$ & $68.3\pm2.7$ \\
Thomson          & $\Gamo=(1-\YP)\,\omega_b$ & $(1-\YP)^2\iB^2\omega_\gamma\,\iM/\iT^2$ & $65.7\pm1.2$ \\
\midrule
\multicolumn{3}{@{}l}{All four (\LCDM)} & $67.78\pm0.42$ \\
\bottomrule
\end{tabular}
\caption{The four routes to $H_0$, in the order of assumptions (i)--(iv). Each assumption links a node anchored by FIRAS to one that is not, and so fixes $\omega_m/\omega_\gamma$ as the function of the invariants shown. Then $h^2=\omega_m/\Om$, with $\Om$ from $\theta_s$. Here $a_\nu=\tfrac78(4/11)^{4/3}$, $N_{\rm eff}=N_{fs}=3.044$, $\YP=\YP^{\rm BBN}(\omega_b)$, and $\Gamo$ is in the units of Table~\ref{tab:invariants}. The last row makes all four assumptions at once (End Matter). $H_0$ is in units of km\,s$^{-1}$\,Mpc$^{-1}$.}
\label{tab:routes}
\end{table}

\textit{Results.---} We constrain the six invariants with the primary CMB power spectra of Planck, ACT DR6 and SPT-3G D1~\cite{Planck:2018vyg,AtacamaCosmologyTelescope:2025blo,SPT-3G:2025bzu}. Adopting a single assumption only fixes the scale, to which the spectra are insensitive, and so does not affect the posterior on the invariants. For convenience, we therefore obtain this posterior in a model in which the pressure-support assumption (ii) is adopted and the other three are relaxed (see End Matter for implementation details). The results are presented in Table~\ref{tab:invariants}, and the full posterior is shown in Fig.~\ref{fig:corner}. Of these, we highlight the $3\%$ constraint on the fraction of pressure-supported matter, $f_{\rm ps} = 0.1580 \pm 0.0047$, which comes from the gravitational signature of pressure support alone, without the additional information that enters in \LCDM\ once the other assumptions, chiefly (i), are also made.

\begin{figure}
  \includegraphics{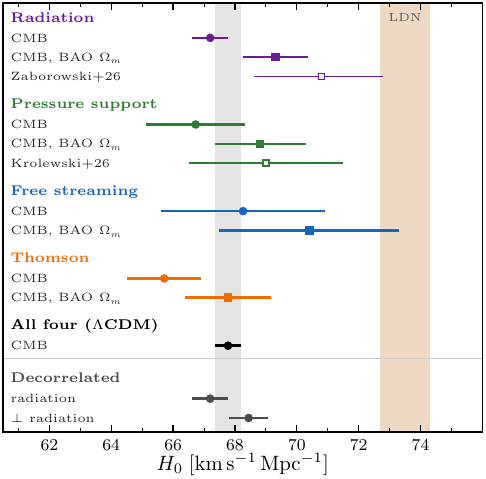}
    \caption{Determinations of $H_0$ grouped by route. Filled circles are the values of Table~\ref{tab:routes}, with $\Om$ from $\theta_s$. Filled squares instead use $\Om$ from DESI DR2 baryon acoustic oscillations (BAO) alone, marginalized over $r_d$~\cite{DESI:DR2}. Open squares are determinations in which DESI DR1 galaxy clustering, in place of the CMB, supplies the route's invariant, with $\Om$ from uncalibrated BAO including $\theta_\star$: Zaborowski et al.~\cite{Zaborowski:2025umc} for $\iM$ and Krolewski et al.~\cite{Krolewski:2025deb} for $\fps$. Below the rule, the four CMB determinations are re-expressed as two uncorrelated ones (see text). The grey band is the \LCDM\ value, which takes all four assumptions at once, and the tan band is the Local Distance Network (LDN)~\cite{H0DN:2025lyy}.}
  \label{fig:forest}
\end{figure}

We obtain $H_0$ along each route by evaluating its expression in Table~\ref{tab:routes} on samples from the posterior, with $\omega_\gamma$ from FIRAS and $\Om$ from $\theta_s$. The four determinations agree (Table~\ref{tab:routes}). Because the routes are evaluated on the same posterior, in which the invariants are themselves correlated, and because they all share $\Om$ and some share other invariants, they are correlated, with coefficients as large as $0.72$. We therefore quantify their agreement by assessing whether the three independent differences between them are consistent with zero. We find that they are, with $\chi^2=3.47$ for three degrees of freedom and a probability to exceed of $0.32$ (Fig.~\ref{fig:corner}). The largest difference, between the Thomson and radiation routes, is $-1.7\sigma$. The four routes, each of which depends on different assumptions and physical effects in their inference of $H_0$, thus give a consistent account of the expansion rate. Taking all four assumptions at once, as is done in \LCDM, gives $H_0=67.78\pm0.42$.\footnote{All quoted values of $H_0$ are in units of km\,s$^{-1}$\,Mpc$^{-1}$.}

We show these determinations as circles in Figure~\ref{fig:forest}. Since every route goes through the late-time edge, replacing the $\Om$ obtained from $\theta_s$ with that determined from DESI DR2 BAO with the sound horizon left free, $0.2975\pm0.0085$~\cite{DESI:DR2}, raises all four by the same $3\%$ (filled squares), to $69.3\pm1.1$ along the radiation route and $68.8\pm1.5$ along the pressure-support route. 

Because the four determinations are correlated, their error bars in Fig.~\ref{fig:forest} cannot be compared directly. We therefore isolate what the other three add by forming the minimum-variance combination of the four that is uncorrelated with the radiation route (last two rows of Fig.~\ref{fig:forest}). It gives $H_0=68.44\pm0.63$, against $67.20\pm0.57$ for the radiation route. The other routes thus contribute the equivalent of an independent determination of nearly the same precision, in agreement with the first at $1.5\sigma$.

On the radiation and pressure-support routes, the corresponding invariant can also be supplied by data other than the CMB (open squares). On the radiation route, $\iM$ can be obtained from the scale of matter--radiation equality imprinted on large-scale structure. This is the basis of the determinations of $H_0$ often described as independent of the sound horizon~\cite{Baxter:2020qlr,Philcox:2020xbv,Farren:2021grl,Philcox:2022sgj,Zaborowski:2024car,GarciaEscudero:2025lef,Lu:2026uph,Zaborowski:2025umc}, which fix the radiation density with the photon temperature and the standard neutrino content, as in assumption (i). We compare with the DESI DR1 analysis of \citet{Zaborowski:2025umc}, chosen so that the data constraining $\iM$ overlap as little as possible with ours and so that $\Om$ comes from uncalibrated BAO, as it does for the filled squares (End Matter). That analysis finds $H_0=70.8^{+2.0}_{-2.2}$, consistent with the filled square.

On the pressure-support route, the only external determinations are, to our knowledge, those of Refs.~\cite{Krolewski:2024jwj,Krolewski:2025deb}. We compare with the more recent and more precise of the two. It measures the fraction of the matter that is pressure supported from its imprint on the DESI DR1 galaxy power spectrum, finding $0.161\pm0.010$, in agreement with $\fps=0.1580\pm0.0047$ from the CMB. With the baryon density from BBN, and $\Om$ from the uncalibrated DESI DR1 BAO, again with the Planck $\theta_\star$ as one further BAO point, this gives $H_0=69.0\pm2.5$, also consistent with the filled square. Both external determinations thus agree with the CMB on the calibration.

\textit{Discussion and conclusions.---} Within \LCDM, the CMB determines $H_0$ in four ways, each transferring the FIRAS calibration to the expansion rate through a different modeling assumption and a different combination of the invariants. We have introduced a network that separates the four, and shown that the values of $H_0$ obtained from each are consistent with one another. For the radiation and pressure-support routes, galaxy clustering measures the route's invariant independently of the CMB, and the resulting $H_0$ agrees with ours when both use the same $\Om$. This is a nontrivial test of the consistency of \LCDM\ with current CMB and large-scale structure data, and the model passes it.

The agreement is itself a strong constraint on resolutions of the Hubble tension. Every route lies below the Local Distance Network value, $73.50\pm0.81$~\cite{H0DN:2025lyy}, so if that value is correct, all four determinations are wrong, and, because they agree, wrong by nearly the same amount. A universe with the higher expansion rate would have to appear, along four routes that rely on different assumptions and different physical effects, as a self-consistent \LCDM\ universe with a lower one.

This could come about in three ways. The first is a misestimated $\Om$, which all routes share. However, even with the lower value from BAO, reaching the LDN value requires a matter density higher by $9$ to $18\%$ along every route. The second is for all four assumptions to fail, and to fail by the same factor. 
This is the scaling-symmetry solution of \citet{Cyr-Racine:2021oal}, in which a mirror world dark sector supplies the energy density needed to rescale the free-fall rates. In their implementation, the helium abundance is left free so that the Thomson rate can be rescaled along with them. However, the abundance required to reach the LDN value is well below measurements of the primordial abundance~\cite{Aver:2026dxv}, and nothing explains why the helium abundance should be lower by just the amount that keeps the Thomson rate in step with the free-fall rates.\footnote{Other ways of rescaling the Thomson rate have been proposed, through a variation of the fine-structure constant~\cite{PhysRevD.107.043529} or of the fine-structure constant together with the electron mass~\cite{Greene:2023cro}. In these, too, the size of the variation must be chosen to match the rescaling of the free-fall rates.} The third is for an ingredient absent from our analysis to bias the inferred invariants along every route. Nothing requires such biases on one route to match those on another, or those in galaxy clustering, which agrees with the CMB where it measures the same invariant. That they should preserve the agreement seems unlikely, and we suspect this is why models that reach the LDN value while fitting the CMB have been hard to find.

If a solution is to be found by modifying the cosmological model, the above considerations suggest organizing proposed solutions not by their ingredients, or by the epoch at which they act, but by which of the invariants their effects on the spectra are degenerate with, and which they can therefore bias. The invariants that a model leaves unbiased then determine which of the other quantities along each route must change if $H_0$ is to be higher. The pressure-support route provides an example. For any model that keeps assumption (ii), $h^2=\omega_b/(\fps\,\Om)$, so a higher $H_0$ requires a higher baryon density, a lower $\fps$, or a lower $\Om$. To the extent that a proposed solution leaves the inference of $\fps$ unchanged, the baryon density must rise in proportion to $h^2$, in conflict with BBN. This may be the origin of the tendency of early-universe solutions toward a high baryon density identified by \citet{Giovanetti:2026aku}.

\begin{acknowledgments}
\textit{Acknowledgments.---} We thank Cara Giovanetti for useful conversations. This work was partially supported by DOE Office of Science award DESC0009999 and the Michael and Ester Vaida Endowed Chair in Cosmology and Astrophysics. The authors acknowledge the High Performance Computing Core Facility at the University of California, Davis, for providing computational resources used in this work. 
\end{acknowledgments}

\bibliography{main}
\clearpage
\section*{End Matter}

\textit{Primordial spectrum.---} The symmetry of Eq.~\eqref{eq:symmetry} also requires the primordial spectrum to transform. In the rescaled universe, the mode $\lambda k$ evolves as the mode $k$ did before the rescaling, so the spectra are unchanged only if $\mathcal{P}_{\mathcal{R}}(k) \to \mathcal{P}_{\mathcal{R}}(k/\lambda)$. For a power-law spectrum with amplitude $A_s$ at fixed pivot $k_p$, this is $A_s \to \lambda^{1-n_s}A_s$ with $n_s$ unchanged~\citep{Zahn:2002rr,Cyr-Racine:2021oal}.

\begin{figure*}
  \includegraphics{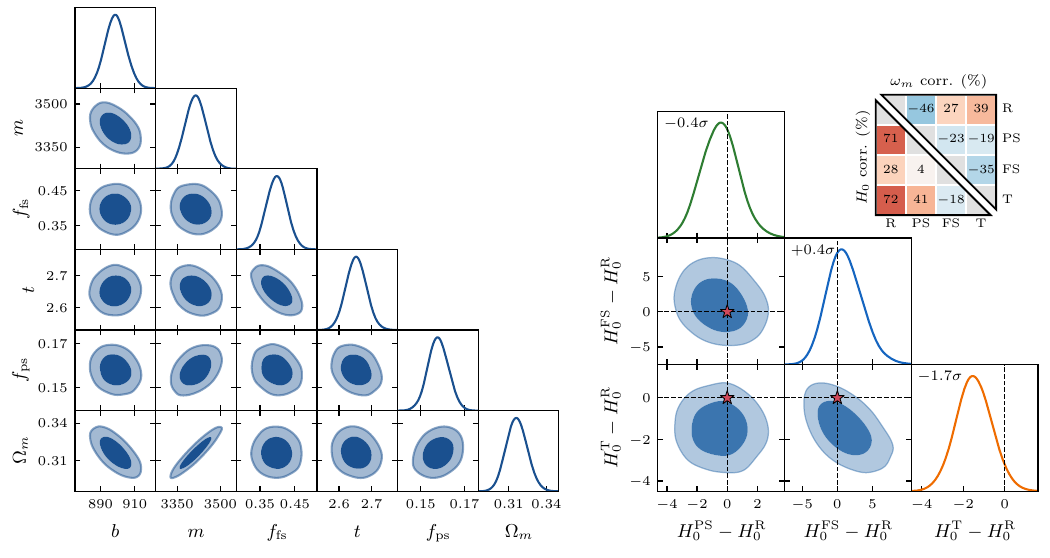}
    \caption{Left: posterior of the invariants from the primary CMB spectra of Planck, ACT DR6 and SPT-3G D1, with $\Om$ shown in place of $\theta_s$. Right: differences between $H_0$ along each route and along the radiation route, in km\,s$^{-1}$\,Mpc$^{-1}$, with $\Om$ from the CMB's $\theta_s$.  The dashed lines and red star mark the origin. Right inset: correlation coefficients between the routes, in percent, for $H_0$ (lower triangle) and for $\omega_m$ (upper triangle).}
  \label{fig:corner}
\end{figure*}

\textit{Sampling.---} We sample $\ln\iB$, $\ln\iM$, $\Rfs$, $\ln\iT$, $\ln\fps$ and $100\,\theta_s$, together with $\ln(10^{10}A_s)$, $n_s$, $\tau$ and four calibration parameters of the likelihoods, with the priors of Table~\ref{tab:S1priors}. We use the Monte Carlo sampler Cobaya~\citep{Torrado:2020dgo,Lewis:2002ah,Lewis:2013hha,Neal:2005sampling}, stopping chains when the Gelman--Rubin statistic satisfies $R-1<0.02$, and we discard the first half of the chain as burn in. For theory predictions, we use the Boltzmann code \textsc{class}~\citep{Blas:2011rf}, emulated with OL\'E~\citep{Gunther:2025xrq}. To test the emulator, we recomputed the posterior with \textsc{class} directly at $3000$ points drawn from the chains and importance-weighted them. No mean moves by more than $0.03\sigma$ and no width by more than $2\%$.

\textsc{class} requires dimensionful inputs, which we obtain by adopting the pressure support assumption ($\omega_b = \omega_{ps}$) and setting the photon density to that of the FIRAS temperature, $T_0=2.7255$\,K~\cite{FIRAS1,Fixsen:1996nj}. From the invariants, we then compute
\begin{equation}
\begin{gathered}
\omega_b=\iB\,\omega_\gamma,\qquad \omega_c=\omega_b\,\frac{1-\fps}{\fps},\qquad \omega_r=\frac{\omega_b+\omega_c}{\iM},\\
a_\nu\omega_\gamma N_{fs}=\Rfs\,\omega_r,\qquad a_\nu\omega_\gamma N_{fld}=(1-\Rfs)\,\omega_r-\omega_\gamma,\\
1-\YP=\frac{\iT\sqrt{\omega_r}}{\omega_b},
\end{gathered}
\label{eq:representative}
\end{equation}
with $a_\nu=\tfrac78(4/11)^{4/3}$. Here $N_{fs}$ and $N_{fld}$ are the effective numbers of free-streaming and fluid-like relativistic species, the latter not counting the photons, so that $N_{fs}+N_{fld}=N_{\rm eff}$. With our assumptions fixing $\xe(a)$, $\sigma_T$ and $m_p$, the helium abundance is the remaining handle available to set $\Gamo$. This model is the ``Mix+$\YP$'' model of \citet{Ge:2022qws}, i.e. \LCDM\ with free $N_{fs}$, $N_{fld}$ and $\YP$.

\begin{table}[t]
\centering
\begin{tabular}{@{}ll@{}}
\toprule
Parameter & Prior \\
\midrule
$\ln \iB$         & $U[6.1847,\ 7.2551]$ \\
$\ln \iM$         & $U[7.1327,\ 9.1327]$ \\
$\Rfs$            & $U[0,\ 1]$ \\
$\ln \iT$         & $U[0.2594,\ 1.6594]$ \\
$\ln \fps$        & $U[-2.5257,\ -1.2040]$ \\
$100\,\theta_s$   & $U[1.03,\ 1.06]$ \\
$\ln(10^{10}A_s)$ & $U[1.61,\ 3.91]$ \\
$n_s$             & $U[0.8,\ 1.2]$ \\
$\tau$            & $U[0.01,\ 0.8]$ \\
$A_{\rm ACT}$     & $\mathcal{N}(1,\ 0.003)$ \\
$p_{\rm ACT}$     & $U[0.9,\ 1.1]$ \\
$T_{\rm cal}$     & $\mathcal{N}(1,\ 0.0036)$ \\
$E_{\rm cal}$     & $U[0.8,\ 1.2]$ \\
\bottomrule
\end{tabular}
\caption{Priors on the sampled parameters. $U$ is uniform and $\mathcal{N}(\mu,\sigma)$ Gaussian. The ranges correspond to $0.012\le\omega_b\le0.035$, $0.08\le\fps\le0.30$, and $\iM$ and $\iT$ within factors of $e$ and $e^{0.7}$ of their values in a fiducial \LCDM\ model. Samples are further restricted to $N_{fld}\ge-1.5$, $0.1<\YP<0.4$ and $N_{\rm eff}<10$, computed from the invariants with Eq.~\eqref{eq:representative}. $A_{\rm ACT}$ and $p_{\rm ACT}$ are ACT DR6 calibration parameters, with the Planck calibration set equal to $A_{\rm ACT}$; $T_{\rm cal}$ and $E_{\rm cal}$ are SPT-3G D1 calibration parameters.}
\label{tab:S1priors}
\end{table}

We treat the additional fluid-like radiation using the dark radiation component of \textsc{class} in its fluid limit, which represents radiation that is tightly self-coupled~ \cite{Cyr-Racine:2015ihg,Archidiacono:2019wdp}. We make one modification to this treatment, which is to allow $N_{fld}<0$. \textsc{class} computes the density of this component from a temperature $T\propto N_{fld}^{1/4}$, which we extend by computing $T$ from $|N_{fld}|$ and assigning the density the sign of $N_{fld}$.

A negative $N_{fld}$ corresponds to a total fluid-like radiation density below that of the photons alone, which is not possible. We introduce it because the spectra constrain $\omega_{fld}$ and $\omega_{ps}$ only through their ratio, the invariant $\omega_{fld}/\omega_{ps}=(1-\Rfs)/(\iM\fps)$. With $\omega_{ps}=\omega_b$, as in our sampling model, $1+a_\nu N_{fld} = (\omega_{fld} / \omega_{ps})/(\omega_\gamma/\omega_b)$, so $N_{fld}<0$ states that $\omega_{fld} / \omega_{ps}$ is smaller than the ratio of photons to baryons. By Eq.~\eqref{eq:symmetry}, the same spectra are produced by a universe in which every rate is larger by the same factor, such that $\omega_{fld}=\omega_\gamma$ and $\omega_{ps}>\omega_b$. That universe contains pressure-supported matter in addition to the baryons, and all of its densities are physical. Requiring $N_{fld}\ge0$ would therefore exclude physical universes of this kind. For reference, requiring $N_{fld}\ge0$ would shift $\Rfs$ by $0.6\sigma$ and raise $H_0$ along the free-streaming route by $1.6$\,km\,s$^{-1}$\,Mpc$^{-1}$.

Finally, the ionization fractions of hydrogen and helium as functions of redshift, not including the reionization contribution, are fixed to those of a fiducial \LCDM\ model computed with HyRec~\cite{PhysRevD.83.043513,Lee:2020obi}. 

\textit{Data.---} At low multipoles we use the Planck 2018 TT \texttt{Commander} and EE \texttt{SimAll} likelihoods~\citep{Planck:2019nip,Planck:2018vyg}. At high multipoles we use the P-ACT combination of \citet{AtacamaCosmologyTelescope:2025blo}, which is the Planck 2018 \texttt{plik\_lite} likelihood restricted to $\ell\le1000$ in TT and $\ell\le600$ in TE and EE, with the ACT DR6 CMB-only likelihood~\citep{AtacamaCosmologyTelescope:2025vnj}, together with the foreground-marginalized SPT-3G D1 likelihood~\citep{SPT-3G:2025bzu} as implemented in \textsc{candl}~\citep{Balkenhol:2024obe}. We importance sampled chains to include CMB lensing reconstruction data \citep{quUnifiedConsistentStructure2025}, and found negligible impact on constraints. As such, we do not report those constraints here.

\textit{Route calculations.---} We evaluate the expressions of Table~\ref{tab:routes} on each sample, with $\omega_\gamma=2.473\times10^{-5}$. For the Thomson route, $\YP^{\rm BBN}$ is the PArthENoPE prediction tabulated in \textsc{class} \citep{2022CoPhC.27108205G}, evaluated at the sample's $\omega_b=\iB\,\omega_\gamma$ and for the standard expansion rate during nucleosynthesis, i.e. with $N_{\rm eff}=3.044$. Evaluating it instead at the radiation density of each sample raises $H_0$ along this route by only $0.3$\,km\,s$^{-1}$\,Mpc$^{-1}$. For the filled squares of Fig.~\ref{fig:forest}, we divide each sample's $\omega_m$ by a value of $\Om$ drawn independently from a chain that fits the DESI DR2 BAO measurements~\cite{DESI:DR2} with $\Om$ and the sound horizon $r_d$ as free parameters, which gives $\Om=0.2975\pm0.0085$.

The right panel of Fig.~\ref{fig:corner} shows $\Delta_i=H_0^{(i)}-H_0^{\rm R}$ for the other three routes. With $\bar\Delta$ and $C_\Delta$ their mean and covariance over the samples, $\chi^2=\bar\Delta^{\rm T}C_\Delta^{-1}\bar\Delta=3.47$, which we compare with a $\chi^2$ distribution with three degrees of freedom. The last two rows of Fig.~\ref{fig:forest} are the radiation route and the combination of the four routes, with weights summing to one, that is uncorrelated with it and has the smallest variance: $2.60\,H_0^{\rm R}-0.55\,H_0^{\rm PS}-0.22\,H_0^{\rm FS}-0.82\,H_0^{\rm T}=68.44\pm0.63$. The inverse-variance weighted mean of the two is the minimum-variance combination of all four routes, $67.76\pm0.43$.

Finally, with (ii) adopted, assumptions (i), (iii) and (iv) are three relations among the invariants: $\iM=\iB/[\fps(1+a_\nu N_{\rm eff})]$, $\Rfs=a_\nu N_{\rm eff}/(1+a_\nu N_{\rm eff})$ and $\YP=\YP^{\rm BBN}(\omega_b)$, with $N_{\rm eff}=3.044$. The last row of Table~\ref{tab:routes} is obtained by approximating the posterior as Gaussian in $(\ln\iB,\ln\iM,\Rfs,\ln\iT,\ln\fps,\ln\Om)$ and analytically conditioning it on these relations.

\textit{External determinations.---} The sound-horizon-independent determinations differ from ours, and from one another, in the data that constrain $\iM$ and that supply $\Om$. We exclude analyses that use CMB lensing, which is reconstructed from the same maps as our spectra. So that any difference reflects $\iM$ and not the late-time edge, we also require that $\Om$ come from uncalibrated BAO and not from supernovae, which favor a higher $\Om$ and give a correspondingly lower $H_0$~\cite{Philcox:2022sgj,Zaborowski:2024car}. The analysis of \citet{Zaborowski:2025umc} meets both requirements: it constrains $\iM$ from the matter--radiation equality scale in the galaxy power spectrum, without using the sound horizon, and $\Om$ from the uncalibrated DESI DR1 BAO together with the Planck $\theta_\star$.

\end{document}